\documentclass[aps,superscriptaddress,showpacs,floatfix,amsmath,amssymb,nofootinbib,preprintnumbers,
twocolumn]{revtex4-1}

\usepackage{graphicx}

\usepackage{epstopdf}

\usepackage{color}

\newcommand{\be}{\begin{equation}}
\newcommand{\ee}{\end{equation}}
\newcommand{\ba}{\begin{eqnarray}}
\newcommand{\ea}{\end{eqnarray}}

\newcommand{\beq}{\begin{equation}}
\newcommand{\eeq}{\end{equation}}
\newcommand{\beqa}{\begin{eqnarray}}
\newcommand{\eeqa}{\end{eqnarray}}

\newcommand{\bea}{\begin{eqnarray}}
\newcommand{\eea}{\end{eqnarray}}

\begin{document}
\title{{Horizon Thermodynamics from Einstein's Equation of State}}
%\title{Horizon Thermodynamics Strikes Back}
%\title{Break Down of Universality of $P-V$ Criticality in Horizon Thermodynamics}

\author{Devin Hansen}
\email{dhansen@perimeterinstitute.ca}
\affiliation{Perimeter Institute, 31 Caroline St. N., Waterloo,
Ontario, N2L 2Y5, Canada}
\affiliation{Department of Physics and Astronomy, University of Waterloo,
Waterloo, Ontario, Canada, N2L 3G1}
\author{David Kubiz\v n\'ak}
\email{dkubiznak@perimeterinstitute.ca}
\affiliation{Perimeter Institute, 31 Caroline St. N., Waterloo,
Ontario, N2L 2Y5, Canada}
\affiliation{Department of Physics and Astronomy, University of Waterloo,
Waterloo, Ontario, Canada, N2L 3G1}
\author{Robert B. Mann}
\email{rbmann@uwaterloo.ca}
\affiliation{Department of Physics and Astronomy, University of Waterloo,
Waterloo, Ontario, Canada, N2L 3G1}
%\affiliation{Perimeter Institute, 31 Caroline St. N., Waterloo,
%Ontario, N2L 2Y5, Canada}

%\date{\today}  % revised version
%\date{\today}
\date{October 10, 2016}

\begin{abstract}
By regarding the Einstein equations as equation(s) of state, we demonstrate that a full cohomogeneity horizon first law can be derived in horizon thermodynamics. In this approach  both the  entropy and the free energy are  derived concepts, while the standard (degenerate) horizon first law is recovered by a Legendre projection from the more general one we derive.
These results readily generalize to higher curvature gravities and establish a way of how to formulate consistent black hole thermodynamics  without conserved charges.
\end{abstract}

\maketitle

%%%%%%%%%%%%%%%%%%%%%%%%%%%%%%%%%%%%%%%%%%%%%%%%%%%%%%%%%
%%%%%%%%%%%%%%%%%%%%%%%%%%%%%%%%%%%%%%%%%%%%%%%%%%%%%%%%%

The discovery that spacetimes with horizons can be well described by thermodynamic laws \cite{Bekenstein:1973ur, Bardeen:1973gs, Hawking:1974sw} has lead to much speculation about the thermodynamic meaning of gravitational field equations \cite{Jacobson:1995ab, Hayward:1997jp, Padmanabhan:2003gd}. Among these, the concept of \emph{horizon thermodynamics} emerged from the discovery that Einstein's equations on the black hole horizon
can be interpreted as a thermodynamic identity \cite{Padmanabhan:2002sha}. First formulated for the spherically symmetric black holes in the Einstein gravity, horizon thermodynamics has since been extended to higher curvature gravities \cite{Paranjape:2006ca, %Kothawala:2009kc, Tian:2010yb,
Sheykhi:2014rka}, time evolving \cite{Kothawala:2007em, Cai:2008mh} and rotating \cite{Kothawala:2007em, Hansen:2016wdg} black hole horizons, or even general null surfaces \cite{Chakraborty:2015hna}.

The key idea of horizon thermodynamics is to realize that the radial Einstein equation\footnote{In this Letter we focus on the most robust formulation of horizon thermodynamics in the presence of spherical symmetry.} when evaluated on the black hole horizon assumes the suggestive form
\be\label{state}
P=P(V,T)\,,
\ee
or in other words  an {\em horizon equation of state},
which comes by making an assumption that the radial component of the stress-energy tensor serves as a thermodynamic pressure, $P=T^r{}_r|_{r_+}$, the temperature is identified with the Hawking temperature, $T=T_H$, and the horizon is assigned a geometric volume $V=V(r_+)$ \cite{Parikh:2005qs, Kubiznak:2016qmn}.

By considering a {\em virtual displacement} of the horizon \cite{Padmanabhan:2002sha}, the horizon equation of state can be rewritten as
a \emph{horizon first law}
\be\label{HFL1}
\delta E=T\delta S-P\delta V\,,
\ee
%\tcr{{\bf The following needs to be modified:}
%where  the black hole entropy $S$ must be specified by other means (e.g. \cite{Iyer:1994ys}), and $E$ is a derived quantity, identified with the black hole {\em quasilocal energy}.}
where $S$ stands for the horizon entropy and $E$ is identified as a quasilocal energy of the black hole.
For example, in  Einstein gravity $E$ turns out to be the Misner--Sharp energy  \cite{Misner:1964je} and the obtained horizon first law \eqref{HFL1} is a special case of the `unified first law' discussed by Hayward \cite{Hayward:1997jp}.  %\tcb{{\bf do we have cases where $E$ is not the Misner--Sharp mass?}}
%\tcr{\bf Why is this important?}

While these results are rather suggestive, there are several issues in this procedure that arise upon further inspection.
%\tcr{{\bf The following needs to be modified:}
% First,  all  relevant thermodynamic quantities must already be known  in order to identify them in the field equations. Namely, $S,T$ and $V$ have to be {\em independently specified} and the only derived quantity is the quasilocal energy $E$.
%Consequently this procedure cannot be used as a way to derive any thermodynamic properties of a spacetime; instead, it serves purely as means to identify a peculiar relationship between the field equations and thermal systems provided the thermodynamic properties of the solution are already known.}
First, in the original derivation, it was unclear which thermodynamic variables were derived and which needed to be independently specified.  The focus was previously on the provocative relation hidden within the Einstein Equations when the appropriate identifications were made.  Consequently this procedure provides no direct algorithmic method to derive thermodynamic properties of a spacetime where appropriate identifications are yet unknown, and has instead been used as means of highlighting the presence of known thermodynamics in the gravitational field equations.

The second issue concerns the restriction to virtual displacements  $\delta r_+$ of the horizon radius.  This renders
the first law \eqref{HFL1} to be of `{\em cohomogeneity-one}', since both $S$ and $V$ are functions only of $r_+$. Indeed
\eqref{HFL1}  could  just as well be written as $\delta E = (TS^\prime + PV^\prime)\delta r_+$, with primes denoting differentiation with respect to $r_+$.  This yields an {\em ambiguity} between `heat' and `work' terms and leads to a `vacuum interpretation' of the first law \eqref{HFL1} \cite{Hansen:2016wdg}.
%as discussed in App.~A in \cite{Hansen:2016wdg}.
%as a `standard vacuum' first law of black hole thermodynamics from a point of view of an observer who can measure the true Hawking temperature,

Here we show that both of the above dilemmas can be avoided.
%by developing the horizon thermodynamics from a new starting point.
The key idea is to vary the horizon equation of state \eqref{state}, treating the pressure $P$ and temperature $T$ as independent thermodynamic quantities. This results in a {\em new horizon first law}
\be\label{HFL2}
\delta G=-S\delta T+V\delta P\,,
\ee
which is manifestly non-degenerate and of cohomogeneity-two. Moreover, upon specifying the volume, pressure, and temperature, the horizon entropy $S$
is now a {\em derived concept} and so is the Gibbs free energy $G$. The standard horizon first law \eqref{HFL1} can be recovered a-posteriori, by applying a degenerate Legendre transformation,
\be\label{Legendre}
E=G+TS-PV\,.
\ee
This new derivation implies that horizon thermodynamics has considerable utility, and provides further evidence that gravitational field equations can indeed be understood as an equation of state.

We begin our discussion by briefly reviewing %generalizing
the traditional cohomogeneity-one
approach to horizon thermodynamics in  four-dimensional Einstein gravity~\cite{Padmanabhan:2002sha}, % to general spherically symmetric spacetimes,
emphasizing which quantities are assumed and which can be obtained as an output. Throughout we employ the units in which $G = c = \hbar = 1$.
Consider a static spherically symmetric black hole spacetime described by the geometry
\bea\label{sss}
	ds^2 &=& -f(r)dt^2 + \frac{dr^2}{g(r)} + r^2d\Omega^2\,,
\eea
with a non-degenerate horizon located at $r=r_+$, determined as the largest positive root of $f(r_+)=0$.
 In what follows we concentrate on the case when $f(r)=g(r)$; the general case $f(r)\neq g(r)$ is conceptually more subtle and will be treated elsewhere \cite{Hansen:InPrep}.
 Assuming minimal coupling to the matter, with the stress energy tensor $T_{ab}$, the radial Einstein equation evaluated on the horizon reads
\be\label{eineq}
8\pi T^r{}_r|_{r_+} = G^r{}_r|_{r_+}= \frac{f'(r_+)}{r_+ }-\frac{1-f(r_+)}{r_+^2}\,,
\ee
where primes denote differentiation with respect to $r$.
Identifying
\be\label{other}
P=T^r{}_r|_{r_+}\,, \quad
T=\frac{ f'(r+) }{4\pi}\,,
\ee
as the respective pressure and temperature yields
\be\label{stateEE}
P= \frac{T}{2r_+ } - \frac{1}{8\pi r_+^2}\,,
\ee
which is the horizon equation of state \eqref{state}. Multiplying this  by $4\pi r_+^2\delta r_+$ then gives
\be\label{HFL1a}
	\frac{\delta r_+}{2}= T\delta S - P\delta V\,,
\ee
which is the horizon first law \eqref{HFL1}, provided we either identify any one of the three quantities
\be\label{S}
V=\frac{4}{3}\pi r_+^3\,, \quad S = \frac{A}{4}=\pi r_+^2\,,   \quad E=\frac{r_+}{2}\,,
\ee
as the volume, entropy, and energy respectively, assuming the latter is a function only of $r_+$.
Identification of the remaining quantities logically follows from
\eqref{HFL1}.  Regardless, the obtained first law \eqref{HFL1}
is cohomogeneity one, as its every term varies solely with $r_+$, and suffers from the ambiguity of defining independent heat and work terms.
However the degree of cohomogeneity in the HFL is a consequence of the procedure chosen and not  intrinsic to horizon thermodynamics itself as we shall now demonstrate.

The identification of the temperature $T$  as in \eqref{other} is via standard arguments in thermal quantum field theory; it does not require any gravitational field equations.  By definition the pressure is identified with the matter stress-energy as in
\eqref{other}. With this information the radial Einstein equation can be rewritten as
\be\label{eosg}
P  =  B(r_+) + C(r_+) T\,,
\ee
where $B$ and $C$ are some known functions of $r_+$
 that in general depend on the theory of gravity under consideration, as does the linearity of the equation of state in the temperature $T$.  Formally varying the generalized equation of state \eqref{eosg}, we obtain
\be\label{eosgvar}
V\delta P =  V \left(B^\prime + C^\prime T \right) \delta r_+    + V C \delta T\,,
\ee
upon multiplication by a function $V(r_+)$ that we shall identify as the volume, assuming all other parameters are fixed.  It is now straightforward to rewrite this equation as
\be\label{HFLG}
V\delta P = S\delta T+\delta G\,,
\ee
where
\bea
G &=&  \int^{r_+}\!\!\!\! dxV(x) B^\prime(x)+ T\int^{r_+}\!\!\! dx V(x) C^\prime(x)
\nonumber\\
&=& PV-ST -\int^{r_+}\!\!\! dxV^\prime(x) B(x)\,,   \label{GSdef} \\
S &=& \int^{r_+}\!\!\! dx V^\prime(x) C(x)\,,     \nonumber
\eea
using the integration by parts.
Since (by postulate) we have identified $T$ with temperature, $P$ with pressure, and $V$ with volume, we therefore conclude that $S$ is the {\em entropy} and $G$ is the {\em Gibbs free energy} of the black hole.  Note that these are {\it derived} quantities from the premises \eqref{other}, and the field equations that yield  \eqref{eosg}, along with the assumption that
the volume (whose explicit form \eqref{other} was not really required up to now) does not depend on $T$.

The relation \eqref{HFLG} for the Gibbs free energy $G=G(P,T)$ is the cohomogeneity-two horizon first law \eqref{HFL2}, where  $P$ and $T$ are  independent quantities.  It is valid for any gravitational theory whose field equations
yield a linear relation between pressure and temperature.
Note that since $G$ depends on the matter content only implicitly (via $P$ and $T$) it characterizes the gravitational theory. This is the origin of recently observed `universality' of the corresponding phase behavior \cite{Hansen:2016ayo}.

%By employing the Euler scaling argument, e.g. \cite{Kastor:2009wy}, the obtained first law is accompanied by the following (four-dimensional) Smarr relation:
%, which in four dimensions reads %\tcr{\bf please check}
%\be
%G=ST-2VP\,.
%\ee

%We can define the {\em horizon enthalpy} by the associated Legendre transformation $H=H(S,P)=G+TS$, and recover
%\be\label{first2}
%\delta H=T\delta S+V\delta P\,,
%\ee
%which is  another non-degenerate horizon first law,
%accompanied by the Smarr relation $H=2TS-2VP\,.$

We can define the {\em horizon enthalpy} by the associated Legendre transformation $H=H(S,P)=G+TS$, and recover
\be\label{first2}
\delta H=T\delta S+V\delta P\,,
\ee
 which is  another non-degenerate horizon first law.  Likewise we  can employ the Euler scaling argument, e.g. \cite{Kastor:2009wy}, to obtain
\be
H=2TS-2VP\,,
\ee
which is the accompanying (four-dimensional) Smarr relation.

We can also make the degenerate Legendre transformation \eqref{Legendre}, whose degeneracy originates in the fact that $S$ and $V$ both being functions of $r_+$ are not independent quantities, and obtain so the `old' cohomogeneity-one horizon first law \eqref{HFL1}.

Specifying to Einstein gravity in four dimensions, it is straightforward to identify $B(r_+)= -(8\pi r_+^2)^{-1}$ and $C(r_+) = 1/(2r_+)$  from \eqref{stateEE}, yielding from \eqref{GSdef}
\be\label{GSresult}
S=\pi r_+^2\,,\quad
G=\frac{r_+}{3}(1-\pi r_+ T)\,,
\ee
using  the geometric definition \eqref{S} of the volume.
This Gibbs free energy was previously derived and its phase diagrams studied in \cite{Ma:2015llh, Hansen:2016ayo};
%\tcb{\bf This is confusing here, it looks like we mean the STU gibbs, but neither of these papers look at STU};
it is understood as $G=G(P,T)$ through the equation of state $r_+=r_+(P,T)$, \eqref{stateEE}. Performing the degenerate Legendre transformation, \eqref{Legendre}, one finds $E=\frac{r_+}{2}$, in accordance with the previous approach.

We emphasize that the derivation of \eqref{HFLG} depends only on the generalized equation of state having the form  \eqref{eosg}.
At no point was it necessary to use the specific form of the volume $V(r_+)$.  Consequently this new approach to horizon thermodynamics readily extends to higher dimensions and higher-curvature gravities. Let us demonstrate this for black holes in Lovelock gravity.

 Lovelock gravity \cite{Lovelock:1971yv} is a geometric higher curvature theory of gravity that can be considered as a natural generalization of Einstein's theory to higher dimensions---it is the unique higher-derivative theory that gives rise to second-order field equations for all metric components.  In $d$ spacetime dimensions, the Lagrangian reads
 \begin{equation}
\mathcal{L}=\frac{1}{16\pi}\sum_{k=0}^{K}\alpha_{k}\mathcal{L}^{\left(k\right)} + \mathcal{L}_{m}\,,
\label{eq:Lagrangian}
\end{equation}
where $K=\lfloor\frac{d-1}{2}\rfloor$ is
the largest integer less than or equal to $\frac{d-1}{2}$, $\mathcal{L}^{\text{\ensuremath{\left(k\right)}}}$ are the $2k$-dimensional Euler densities, given by $\mathcal{L}^{\left(k\right)}=\frac{1}{2^{k}}\,\delta_{c_{1}d_{1}\ldots c_{k}d_{k}}^{a_{1}b_{1}\ldots a_{k}b_{k}}R_{a_{1}b_{1}}^{\quad c_{1}d_{1}}\ldots R_{a_{k}b_{k}}^{\quad c_{k}d_{k}}\,,$
with the  `generalized Kronecker delta function' $\delta_{c_{1}d_{1}\ldots c_{k}d_{k}}^{a_{1}b_{1}\ldots a_{k}b_{k}}$  totally antisymmetric in both sets of indices, $R_{a_{k}b_{k}}^{\quad c_{k}d_{k}}$ is the Riemann tensor,
and the $\alpha_{\left(k\right)}$ are the  Lovelock coupling constants. In what follows we identify the cosmological constant $\Lambda=-\alpha_{0}/2$, set $\alpha_{1} = 1$ to remain consistent with general relativity, and assume minimal coupling to the matter, described by the matter Lagrangian $\mathcal{L}_{m}$.

Let us consider a static spherically symmetric black hole, allowing now for horizons of various topologies:
\be
ds^2=-f(r)dt^2+ \frac{dr^2}{f(r)}+r^2 h_{ij}dx^idx^j\,.
\ee
Here, $h_{ij}$ ($i, j=2, \ldots, d-1$)  stands for the line element of a $\left( d-2 \right)$-dimensional  space of constant curvature $\sigma(d-2)(d-3)$, with  $\sigma=+1,0,-1$ for spherical, flat, and hyperbolic
geometries respectively of finite  volume  $\Sigma_{d-2}$.

Following \cite{Hansen:2016ayo}, let us include the contribution of the cosmological constant (if present) to the matter part, replacing the definitions in \eqref{other} by
\be
P=T^r{}_r|_{r_+}-\frac{\Lambda}{8\pi} \qquad
T=\frac{ f'(r+) }{4\pi}\,.
\ee
The radial Lovelock equation evaluated on the horizon rewrites as the horizon equation of state, which again assumes the form \eqref{eosg}, where now \cite{Hansen:2016ayo}
\be\label{BCLovelock}
B(r_+)=\sum_{k=1}^K\alpha_k B_k(r_+)\,,\quad
C(r_+)=\sum_{k=1}^K\alpha_k C_k(r_+)\,,
\ee
and
\ba
B_k(r_+ ) &=& -\frac{(d-2k-1)(d-2)!}{16\pi(d-2k-1)!}\left(\frac{\sigma}{ r^2_+}\right)^{k}\,,\nonumber\\
C_k(r_+) &=& {\frac{1}{4 r_+}\frac{k(d-2)!}{(d-2k-1)!}\left(\frac{\sigma}{r^2_+}\right)^{k-1}}\,.	
\ea
Identifying the volume $V$ with the black hole geometric volume \cite{Kubiznak:2016qmn}
\be
V=\frac{\Sigma_{d-2}}{d-1}r_+^{d-1}\,,
\ee
the
formulae \eqref{GSdef} imply
\ba
S&=&\frac{\Sigma_{d-2}}{4}\sum_{k=1}^K\alpha_k\frac{(d-2)!}{(d-2k-1)!}\frac{k\sigma^{k-1}}{d-2k}r^{d-2k}_+\,,\\
G&=&\frac{\Sigma_{d-2}}{d-1}\sum_{k=1}^K\frac{k\alpha_k(d-2)!}{4 (d-2k-1)!} r^{d-2}_+ \left(\frac{\sigma}{ r^2_+ }\right)^{k-1}\times\nonumber\\
&&\quad\times \Bigl[\frac{\sigma(1-\delta_{d,2k+1})}{2\pi  r_+}-\frac{(2k-1)}{d-2k}T\Bigr]\,,
\ea
%\tcr{\bf can the latter formula be written without $\delta$?}
and we recover the horizon first law \eqref{HFL2}.
Note that the derived $S$ is a non-trivial generalization of entropy for the Lovelock black holes \cite{Iyer:1994ys}.
By performing the degenerate Legendre transformation \eqref{Legendre}, we obtain
\be
E=\frac{\Sigma_{d-2}}{16\pi}\sum_{k=1}^K\alpha_k\frac{\sigma^k(d-2)!}{(d-2k-1)!} r^{d-2k-1}_+\,,
\ee
which is the generalized Misner--Sharp energy \cite{Maeda:2011ii}. The degenerate horizon first law \eqref{HFL1}   can  therefore be understood as a special case of the `unified first law' \cite{Maeda:2011ii}.

As argued in \cite{Kastor:2010gq}, to obtain a consistent Smarr relation, the first law has to be extended to contain variations of the Lovelock coupling constants. This is easily achieved in our approach. Namely, starting again from the horizon equation of state \eqref{eosg} with \eqref{BCLovelock}, we can also vary the Lovelock couplings $\alpha_k$ ($k=2,\dots, K$), thereby obtaining a generalized horizon first law
\be\label{HFL3}
\delta G=-S\delta T+V\delta P+\sum_{k=2}^{K}\Psi^k \delta \alpha_k\,,
\ee
where $G$ and $S$, \eqref{GSdef}, are given above and
\ba
\label{eqn:psi}
\Psi^k&=&({\cal C}_k-VC_k)T+{\cal B}_k-VB_k\\
&=&\frac{\Sigma_{d-2}(d-2)!\sigma^{k-1}}{16\pi (d-2k-1)!}r_+^{d-2k}
\Bigl[\frac{\sigma(1\!-\!\delta_{d,2k+1})}{r_+}-\frac{4\pi kT}{d\!-\!2k}\Bigr]\nonumber
\ea
are the conjugate potentials to variable $\alpha_k$; quantities ${\cal C}_k$ were determined from ${\cal C}=\sum_{k=1}^K\alpha_k{\cal C}_k$ and similar for quantities ${\cal B}_k$.
%\tcr{\bf any comments on "vacuum character of $\Psi$'s?} \
Note that by construction the potentials $\Psi^k$ depend on matter only implicitly through the temperature $T$.  Because of this, Eq.~\eqref{eqn:psi} corresponds to the vacuum values of the potentials, c.f. Eq.~(2.23) in \cite{Frassino:2014pha}.
The obtained horizon first law \eqref{HFL3} is obviously of cohomogenity-$(K+1)$.
It is now easy to verify that one obtains the following Smarr relation:
\be
(d-3)H=(d-2)TS-2VP+\sum_{k=2}^K2(k-1)\Psi^k\alpha_k
\ee
for the enthalpy,
completing the horizon thermodynamic description of Lovelock black holes.

To summarize, we have shown that it is possible to re-write the Einstein equations as a manifestly cohomogeneity-2 thermodynamic relation. This solves the problem of the apparent cohomogeneity-1 first law present in horizon thermodynamics since its conception.  We have shown that this process also reduces the number of assumptions needed.  In fact, by specifying the horizon temperature, pressure, and volume, our procedure allows one to {\it derive} the horizon entropy up to an arbitrary constant as well as the free-energy.  Horizon thermodynamics is therefore of equal conceptual power to the extended phase space picture for black hole thermodynamics, e.g. \cite{Kubiznak:2016qmn}.  In contrast to this latter picture, horizon thermodynamics makes no use of conserved quantities, but rather deals only with parameters defined at the horizon.

%on a similar playing field   It shows that horizon thermodynamics can be used to `\emph{derive}' thermodynamics, rather than being simply a clever identification game played on the field equations.

One of the caveats of this approach is that it requires the equation of state, i.e. the field equations, to be written as a linear function of the Hawking temperature $T$.  While this is certainly true for  Einstein and Lovelock gravity, it would be interesting to categorize what types of gravitational theories permit the construction presented here.  It seems reasonable to guess that these theories may simply be in the class of \textit{quasi-topological gravity}~\cite{Myers:2010ru}, as this is a general class of theories that produce at most second order field equations, and whose radial field equation is linear in first derivatives of the metric function on the horizon.  While quasi-topological gravity certainly fits this paradigm \cite{Sheykhi:2014rka}, it is possible that there are other theories whose spherically symmetric black hole solutions can be treated with this procedure.

Another venue for future studies is to extend the present approach to the rotating black hole spacetimes, picking up the threads on the recent progress in \cite{Hansen:2016wdg}.
% \tcb{This approach also seems to extend non-trivially to more general spherically symmetric spacetimes~\cite{Hansen:InPrep}}.

%\tcb{It seems that the calculation presented here can be promoted to rotating black holes, at least in the case of 4-D Einstein gravity, using the ``surface-tenstion" approach of~\cite{Hansen:2016wdg}.  However, more work needs to be done on understanding and motivating generalizations to more complicated metric ansatz and higher curvature theories.}  A possible direction for future study would be to repeat the calculation presented here in such a case and compare to the results obtained in~\cite{Hansen:2016wdg}.

\section{Acknowledgements}
This research was supported in part by Perimeter Institute for Theoretical Physics and by the Natural Sciences and Engineering Research Council of Canada.
Research at Perimeter Institute is supported by the Government of Canada through the Department of Innovation, Science and Economic Development Canada and by the Province of Ontario through the Ministry of Research, Innovation and Science.
\bigskip

%\bibliography{master}
%\bibliographystyle{JHEP}

\providecommand{\href}[2]{#2}\begingroup\raggedright\endgroup

\end{document}